\documentclass[twocolumn,amsmath,amssymb]{revtex4}
\usepackage{graphicx}
\usepackage{dcolumn}
\usepackage{bm}

\begin{document}

\title{Note on description of excited hadrons with fixed angular momentum}

\author{S. S. Afonin}
 \altaffiliation[Also at]{ V. A. Fock
Department of Theoretical Physics, St. Petersburg State
University, 1 ul. Ulyanovskaya, 198504, Russia.}
 \affiliation{Institute for Theoretical Physics II, Ruhr University Bochum,
150 Universit\"{a}tsstrasse, 44780 Bochum, Germany.}


\begin{abstract}
In this Note, we make comments on the recent criticism of  the
description of highly excited hadrons in terms of fixed angular
momentum $L$ and quantify the relation $M^2\sim L+n+S/2$ ($n$ is the
radial quantum number and $S$ is the total quark spin) for masses
of light mesons.
\end{abstract}

\maketitle

\begin{figure*}[tb]
\vspace{-5 cm}
\hspace{-2 cm}
\includegraphics[scale=1.0]{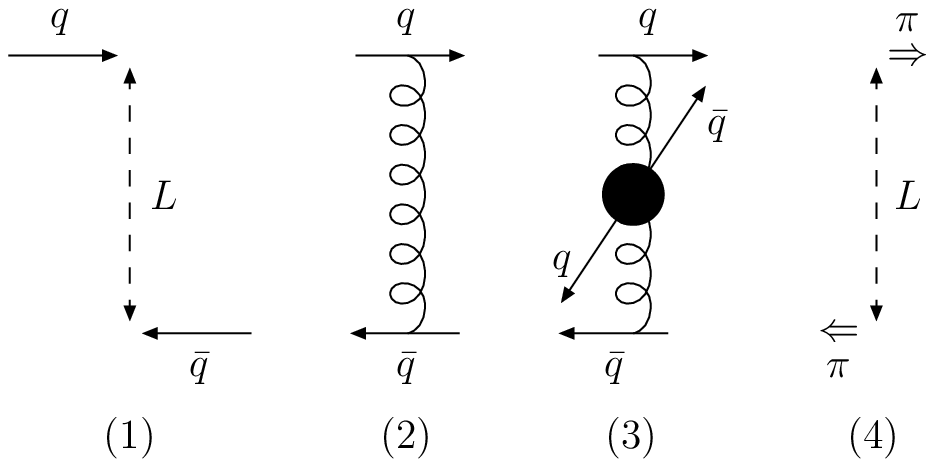}
\vspace{-20cm}
\caption{The formation and decay of excited meson (see the text). The one-gluon
exchange at stage~(2) is the simplest interaction, under this plot we mean all
possible interactions. The same is implied at stage~(3).}
\end{figure*}
In the recent Comment~\cite{gl}, it was attempted to prove that a consistent
holographic description of excited hadrons with fixed angular
momentum $L$ is not possible. This statement is not specific for
the holographic description only, it has a generic character: As
the highly excited hadrons composed of light quarks represent
ultrarelativistic systems, the internal angular momentum $L$
and the total quark spin $S$ are not
conserved separately, hence, any descriptions of excited
hadrons based on fixed $L$ and $S$ (say, the potential models with valence
quarks) are not consistent with the Lorentz invariance.
Since this claim keeps regularly
appearing in the literature and on the conferences, a clear-cut
explanation is called for as to the physical reasons invalidating
such a logic. In part, this question has been already addressed in
the Response~\cite{br} to the Comment above, in the first part of
the present Note we will provide alternative arguments in favor of
the description of excited hadrons at fixed $L$ and discuss the
physical sense of the relation
\begin{equation}
\label{1}
M^2\sim L+n,
\end{equation}
($n$ is the radial quantum number) for meson masses.

The excited hadron is not just a relativistic quantum system, it
is observed as a complicated phenomenon that includes $in$-- and
$out$--states and the description of this phenomenon is
inseparable from the $in$-- and $out$--states, in other words, it
must be clearly defined how it was produced and how it decayed.
This point seems to be misunderstood when making the claim in
question. Let us consider the simplest example --- the creation of
excited light meson and its decay into two pions in the center of
mass system. The whole process can be divided into four stages
which are displayed in FIG.~1. To create a highly excited meson
one needs highly energetic quark $q$ and antiquark $\bar{q}$. Such
a $q\bar{q}$ pair can be produced, for instance, by a
proton-antiproton collision. Due to the asymptotic freedom, the
energetic quarks fly like almost free and well localized objects,
hence, they have a well-defined relative angular momentum $L$ and
intrinsic spin. This first stage of the process represents the
$in$--state. Then the quark and antiquark interact forming a kind
of correlated system (stage~(2)) --- the meson resonance. Strictly
speaking, we can regard this effect as a resonance only if the
correlation of the quark $q$ with the given antiquark $\bar{q}$ is
stronger than with other surrounding quarks and antiquarks. At
stage~(3), the confinement interaction creates an additional
quark-antiquark pair from the vacuum and finally (stage~(4)) we
observe two pions --- the $out$--state. Since the pions are
$S$-wave states (we know this from the fact that they are extensively
produced in the point-like process like the
$e^+e^-$--annihilation) the relative angular momentum of pions is
equal to $L$ due to the angular momentum conservation. {\it $L$ is not
a conserved quantum number at stages~(2) and~(3), but this is
obviously not the case for the $in$-- and $out$--states}. Measuring
experimentally the $L$ of produced pions we find $L$ of initial $q\bar{q}$
pair. This very $L$ enters the relation~\eqref{1}.

The relation~\eqref{1} follows from the analysis of available
experimental data~\cite{af}. First of all, it tells us about the
quantum nature of meson resonances --- the $q\bar{q}$ pair
resonates only at integer values of relative $L$ and at certain
discrete values of $q$ and $\bar{q}$ momentum (dictated by the
integer value of $n$). Second, a large degeneracy takes place due
to the existence of principal quantum number $N\sim L+n$ like in
the hydrogen atom~\cite{af2}. Thus, a theoretical challenge is to
explain the relation~\eqref{1} and to find possible corrections. In
this regard, the potential (or in some sense "quasipotential")
models as well as other models describing the light mesons as
bound states represent a possible line of research, one should
only keep in mind that $L$ and $n$ should not be understood
literally in this case (one deals with models, not with a theory),
they are rather a convenient tool for the description and
classification of resonances.

If the meson spin $J$ is somehow established experimentally, we
can restore the total quark spin $S$ of initial $q\bar{q}$ pair
following the quantum mechanical rule $J=L+S$, $S=0,1$. The
Comment~\cite{br} contains an interesting proposal that $S$
determines the intercept in the relation~\eqref{1}, $M^2\sim
L+n+S/2$. We have performed a fit of experimental data used
in~\cite{af} exploring this suggestion, the result is (in GeV$^2$)
\begin{equation}
\label{2}
S=0: \qquad M^2 = 1.22(L+1.04n+0.22),
\end{equation}
\begin{equation}
\label{3}
S=1: \qquad M^2 = 1.12(L+0.93n+0.63).
\end{equation}
The estimated errors are within 10\% for each constant. If we do
not include the $\pi$-meson the fit~\eqref{2} yields
$M^2 = 1.17(L+1.04n+0.36)$. The relations~\eqref{2} and~\eqref{3}
are to be compared with the original fit~\cite{af} that did not
make separation between the $S=0$ and $S=1$ states,
\begin{equation}
\label{4}
M^2 = 1.10(L+n+0.62).
\end{equation}
Thus, the suggestion above seems to agree qualitatively with the
available experimental data.

In the second part of the present Note, we raise an objection to a
claim in~\cite{gl} stating that the ultraviolet matching condition
used in the holographic models is inconsistent with the chiral
symmetry. The primary theoretical objects in the holographic
approach are the correlation functions, therefore the situation
here is in one-to-one correspondence with the QCD sum rules based
on the Operator Product Expansion --- it is well known that the
manifest chiral invariance of the correlation functions in the
ultraviolet does not imply, generally speaking, the asymptotic
chiral invariance for the physical states saturating the
correlation functions (unless some additional assumptions are
used, see, e.g.,~\cite{af3}). This point can be easily seen
qualitatively: Saturating, for instance, a two-point correlator of
some quark operator by physical mesons with appropriate
quantum numbers (we neglect the decay width and consider the Euclidean
space),
\begin{equation}
\label{5}
\Pi(Q^2)\sim\sum_i\frac{Z_i}{Q^2+m_i^2},
\end{equation}
we have the residues $Z_i$ in the numerators. The physical sense
of quantities $Z_i$ is the probability of creation of the
corresponding meson at the origin. It is quite evident that,
generally speaking, these probabilities strongly depend on $L$
of initial $q\bar{q}$ pair. Since the orbital momenta confronting
the physical states related by the chiral transformations are
different (because of opposite parities of chiral partners,
$P=(-1)^{L+1}$) and not restricted by the chiral symmetry, the
ultraviolet constraint for the correlators related by the chiral
symmetry (let be $\Pi_+$ and $\Pi_-$),
\begin{equation}
\label{6}
\Pi_+(Q^2)-\Pi_-(Q^2)\rightarrow 0,\quad Q^2\rightarrow\infty,
\end{equation}
does not necessary imply for the masses of physical states to
follow the pattern of degeneracy predicted by the chiral
multiplets. The chiral basis seems to be useful for the
classification of correlation functions, but its use for the
physical resonances is not convincingly justified neither
theoretically nor experimentally, instead the standard nonrelativistic
$^{2S+1}L_J$ basis turns out to be more natural and convenient for
the spectroscopy of excited light mesons.

Finally, we would indicate the following delicate point. There
exists an important theoretical case when the orbital momentum
indeed cannot be confronted with the resonances in a
straightforward manner --- the planar limit of QCD,
$N_c\rightarrow\infty$: The mesons do not decay in that limit,
hence, $L$ cannot be deduced experimentally. On the other hand, it
is rather obscure how to make an imaginary experiment on meson
production in the planar world. The point is that the typical
holographic models deal with precisely this limit, so the notion
of $L$ in this case needs some additional clarification. A
possible way out may consist also in effective account for finite
$N_c$ by constructing bottom-up AdS/QCD models describing a final
set of discrete resonances~\cite{af4}.

The author thanks Stan Brodsky and Guy de T\'{e}ramond for
their stimulation to express his insight in the written form.
The research is supported by the Alexander von Humboldt Foundation
and by RFBR, grant 09-02-00073-a.

\end{document}